# Switching spin filling sequence in a bilayer graphene quantum dot through trigonal warping

Guo-Quan Qin[1,2,3], Fang-Ming Jing[1,2,3], Tian-Yue Hao[1,3], Shun-Li Jiang[1,3], Zhuo-Zhi Zhang[1,2,3], Gang Cao[1,3], Xiang-Xiang Song[1,2,3]*, and Guo-Ping Guo[1,3,4]*

1. CAS Key Laboratory of Quantum Information, University of Science and Technology of China, Hefei, Anhui 230026, China
2. Suzhou Institute for Advanced Research, University of Science and Technology of China, Suzhou, Jiangsu 215123, China
3. CAS Center for Excellence in Quantum Information and Quantum Physics, University of Science and Technology of China, Hefei, Anhui 230026, China
4. Origin Quantum Computing Company Limited, Hefei, Anhui 230088, China

* Author to whom correspondence should be addressed:
X.-X. S. (songxx90@ustc.edu.cn) or G.-P. G. (gpguo@ustc.edu.cn)

**Abstract**

We demonstrate a switchable electron shell structure in a bilayer graphene quantum dot by manipulating the trigonal warping effect upon electrical gating. Under a small perpendicular electric field, the lowest s-shell is sequentially filled with two spin-up and two spin-down electrons of opposite valleys. When increasing the electric field, an additional three-fold minivalley degeneracy is generated so that the s-shell can be filled with 12 electrons with the first/last 6 electrons having the same spin polarization. The switched spin filling sequence demonstrates the possibility of using the trigonal warping effect to electrically access and manipulate the spin degree of freedom in bilayer graphene.

Bilayer graphene (BLG) has attracted significant research interest since a tunable band gap can be generated by introducing interlayer asymmetry using an out-of-plane electric field [1-4]. Meanwhile, the skew interlayer coupling between the two carbon layers gives rise to triangular distortions of iso-energetic lines near $K/K'$ points in its Brillouin zone, known as trigonal warping [5]. The combined influence of these two effects makes each band extremum at $K/K'$ points evolve into three local extrema, exhibiting an additional three-fold minivalley degeneracy [6-12]. This band topology plays an important role in determining the behaviors of charge carriers in BLG [8,9,13,14]. For example, a crossover from three-fold degenerated to non-degenerated Landau level spectrum has been observed, indicating the presence of a Lifshitz transition [9,15]. More interestingly, a cascade of correlated phases as well as superconductivity has been revealed in trigonally warped BLG, particularly near Lifshitz transitions or van Hove singularities [6,7,16,17]. These findings demonstrate a promising playground of trigonally warped BLG for studying exotic electronic phases.

To access the three-fold minivalley induced by the trigonal warping, the carrier density needs to be kept low so that the Fermi level is close to the band edge [6]. Therefore, gate-defined quantum dot (QD) devices [18,19] stand out, since their charge carriers can be depleted out of the dot individually. This makes it possible that only few electrons/holes are confined in an area of ~50 nm × 50 nm, corresponding to a charge density as low as ~$10^{10}$ cm$^{-2}$. Moreover, the excellent electrical tunability of QDs offers a powerful knob for manipulating electrons' quantum degrees of freedom (e.g., charge [20-22], spin [23,24], valley [25,26]) at the single-particle level. Recently, high-quality BLG-based QD devices have been achieved [27-31], enabling mapping spin-valley states [32-36], switching Pauli spin/valley blockade [37-40], and extracting spin/valley lifetime [26,41-45]. In particular, energy costs of sequentially loading electrons into BLG-based QDs have been investigated [27,46,47]. When electrons are filled to a higher d-shell, an additional three-fold degeneracy evolves, which hints at an increased influence of minivalleys [46]. It is curious, while remaining unexplored, whether these minivalleys interplay with the two-fold spin/valley degrees of freedom so that they can serve as an electrically-accessible handle to influence the spin and valley magnetic moments.

In this work, we demonstrate a switchable electron filling structure in a BLG-based QD. We mainly focus on the lowest s-shell, which is orbitally non-degenerated, to better reveal the role of the minivalley degeneracy. Under a small perpendicular electric field, the s-shell can be filled with 4 electrons. When increasing the electric field, a more pronounced trigonal warping effect is triggered. The induced three-fold minivalley degeneracy switches the structure of the s-shell so that 12 electrons can be filled. More importantly, based on the switchable shell structure, we are able to map the spin/valley filling sequence in the absence/presence of the minivalley degeneracy to investigate their interplays in between. We demonstrate that the spin filling sequence of the first 12 electrons can be electrically changed from "2+2+4+4" (filling two spin-up, two spin-down, four spin-up, and four spin-down electrons sequentially) to "6+6" (filling six spin-up and six spin-down electrons sequentially), showing the possibility of switching the minivalley degree of freedom to electrically manipulate the spin degree of freedom. Our highly-tunable QD device provides a promising platform for generating high-spin states and for exploring the flavor SU(3) symmetry in solid-state systems [48]. Moreover, the capabilities of controlling the trigonal warping effect and mapping spin/valley sequence using QDs offer an opportunity to investigate the exotic electronic phases in the trigonally warped BLG at the single-particle level.

Figure 1(a) illustrates the device schematic investigated in this work. The detailed fabrication process can be found in Section S1 in Supplemental Material [49]. We electrically characterize the device at the temperature of ~20 mK. Three finger gates (FG1, FG2, and FG3) are responsible for manipulating the electrostatic potential in the conducting channel defined by opposite voltages applied to the metal back and split gates [27]. Here the conducting channel is kept to be of p-type, and an n-type single QD is defined underneath FG2 using natural p-n junctions as tunnel barriers [27,30,37] (see Section S2 in Supplemental Material [49]).

As widely varying the voltage applied to FG2, first 38 Coulomb diamonds are clearly resolved (see Fig. 1(b)). The shell structure of adding extra electrons into the dot can be extracted. For example, Fig. 1(c) zooms in the regime of 3rd, 4th, and 5th diamonds. According to the constant interaction model [18], the height of the 4th diamond corresponds to the energy cost $E_{add}(4)$ for adding an extra electron into the dot when it already contains 4 electrons (see Section S3 in Supplemental Material [49] for more details). Note that $E_{add}(4)$ is larger than $E_{add}(5)$, as well as $E_{add}(3)$. This is because the two-fold spin and two-fold valley degeneracies of BLG allow filling 4 electrons into the s-shell. While adding the 5th electron, a larger amount of energy is needed to overcome the first shell energy (labeled as $\Delta E_{sh1}$). Therefore, we can use $E_{add}(4) - E_{add}(5)$ to estimate $\Delta E_{sh1} = 0.53$ meV [53]. Meanwhile, the excited state spectrum is also resolved in the Coulomb diamond measurement (see the red arrows in Fig. 1(c)). The excited state energy of the 4th electron can be extracted as $E_{4es}^{+} = 0.50$ meV ($E_{4es}^{-} = 0.52$ meV) at the positive (negative) $V_{SD}$ branch from Fig. 1(d). We find $E_{4es} = (E_{4es}^{+} + E_{4es}^{-})/2 = 0.51$ meV is close to $\Delta E_{sh1}$. The slight difference probably originates from the variation in the component of the charging energy in determining $\Delta E_{sh1}$ [49]. This influence is expected to be more pronounced at the few-electron regime, where the dot size can be strongly modulated even if one extra electron is loaded. Nevertheless, this result indicates that, in addition to being filled to the s-shell, the 4th electron can also be filled to an excited state in the second p-shell with an energy separation of $E_{4es} \approx \Delta E_{sh1}$ [54]. Similar results are found for the 12th, 24th, and 36th Coulomb diamonds, where their addition energies exhibit local maximums as well (see Section S4 in Supplemental Material [49]). Our observation demonstrates the shell structure of filling 4, 8, 12, 12 electrons to s-, p-, d-, f-shells, respectively. Except for the lowest s-shell, higher shells exhibit additional degeneracies, beyond the four-fold spin and valley degeneracies. This may originate from the confinement-induced orbital degeneracy. For example, a two-dimensional isotropic parabolic confinement leads to an *n*-fold orbital degeneracy for the *n*-th shell [55], known as the Fock-Darwin shell structure [18]. Recently, studies on circular BLG-based QDs suggest that the filling of 12 electrons in the d-shell is related to the emerged influence of the three-fold minivalley degeneracy induced by the trigonal warping [46,56]. In this work, to avoid the difficulties in identifying the origin of the additional degeneracy for higher shells, we focus on the orbitally non-degenerated s-shell to better reveal the role of the trigonal warping.

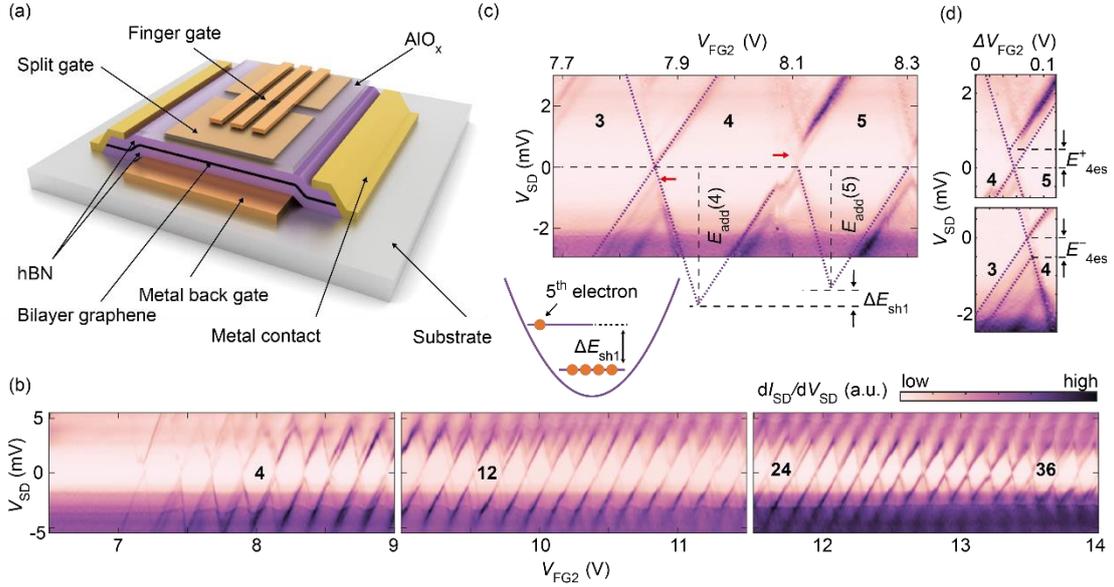

**Fig. 1** (a) Schematic of the device. The heterostructure, consisting of a BLG flake (black) encapsulated in two hBN flakes (purple), is placed on a metal back gate (BG, dark orange). The split gates (SGs, light orange) and finger gates (FGs, light orange) are separated by an insulating aluminum oxide layer. The metal contacts (yellow) are used to detect the transport current. (b) Finite bias spectroscopy measurement for the device. The numbers $N$ indicate the electron occupation in the regions of the Coulomb blockade. (c) Zoom-in of the 3rd, 4th, and 5th Coulomb diamonds, from which the first shell energy $\Delta E_{\text{sh1}}$ is extracted. Red arrows indicate the transition lines corresponding to the excited states. The bottom left schematic illustrates the shell structure of the first 5 electrons. (d) Extracting the excited state energy of the 4th electron from the positive (upper panel) and the negative $V_{\text{SD}}$ branches (lower panel), respectively.

It has been previously demonstrated that the three local minivalleys near the band edge in BLG can be deepened to the order of a few meV upon increasing the perpendicular electric field [15,56]. Such a value is larger than the extracted shell energy $\Delta E_{\text{sh1}}$. Therefore, if a large electric field is applied, the influence of the trigonal warping can be triggered even at the lowest s-shell. Figure 2(a) shows the first 12 Coulomb peaks measured at a small electric field of 0.36 V/nm (determined by the COMSOL simulation). We extract the peak spacing $\Delta V_{\text{peak}}$ and the addition energy $E_{add}$ as a function of electron number $N$ in the dot, as shown in Figs. 2(b) and 2(c) respectively. Similar maximums at $N = 4$ and $N = 12$ are observed. This is consistent with the Fock-Darwin shell structure which two-fold spin and two-fold valley degeneracies are involved. Namely, the lowest s-shell is occupied by four electrons with opposite spins and valleys, while the p-shell has two orbitals. Each orbital can be filled with four electrons as well. The shell structure is illustrated in Fig. 2(d), which is referred to as the case of "four-fold degeneracy" for simplicity. Interestingly, when increasing the electric field to 0.53 V/nm, the maximums in $\Delta V_{peak}$ and $E_{add}$ at $N = 4$ disappear (see Figs. 2(f) and 2(g)). This means all of the first 12 electrons are filled to the s-shell. Since the s-shell is orbitally non-degenerated, this clearly suggests the pronounced influence of the three-fold minivalley degeneracy. The corresponding shell structure is illustrated in Fig. 2(h), which is referred to as the case of "twelve-fold degeneracy".

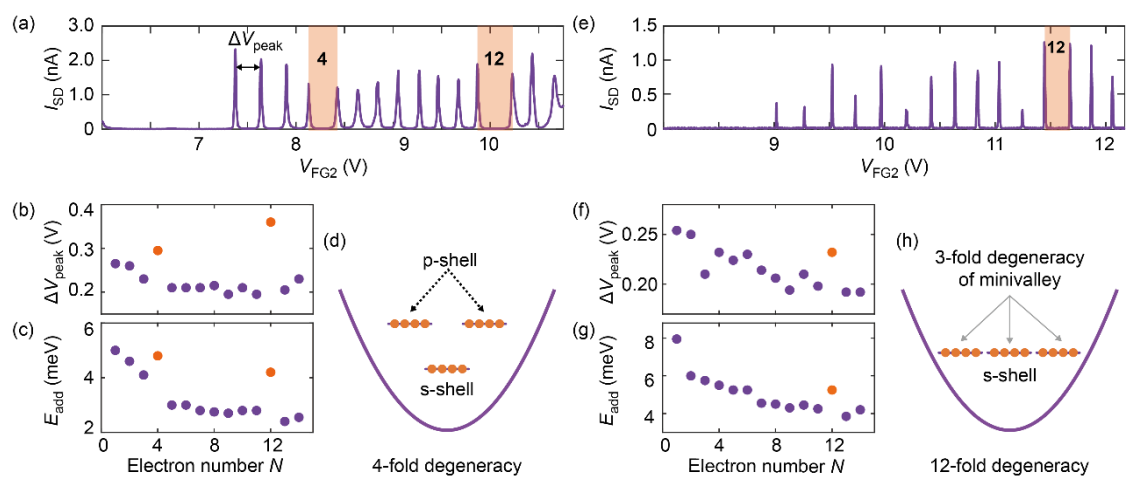

**Fig. 2** Current through the dot as a function of $V_{FG2}$ at the perpendicular electric field of (a) 0.36 V/nm and (e) 0.53 V/nm, respectively. (b) Peak spacing $\Delta V_{peak}$ and (c) addition energy $E_{add}$ as a function of electron number $N$ in the dot, which are extracted from (a). (d) Schematic shell structure for the case of four-fold degeneracy, corresponding to (a). (f) $\Delta V_{peak}$ and (g) $E_{add}$ as a function of $N$, which are extracted from (e). (h) Schematic shell structure for the case of twelve-fold degeneracy, corresponding to (e). Electron numbers corresponding to full-shell are highlighted in orange in (b-c) and (f-g), respectively.

To further understand the switch from four-fold to twelve-fold degeneracy in the s-shell, we investigate the spin/valley filling sequence under external magnetic fields. The device is re-bonded and cooled to ~20 mK in another refrigerator equipped with a magnet. A similar switch upon increasing the electric field is found (see Section S5 in Supplemental Material [49]).

Figure 3(a) shows the evolution of the first 12 Coulomb peaks as a function of in-plane magnetic field $B_\parallel$ for the case of four-fold degeneracy (at a perpendicular electric field of 0.47 V/nm). The shift of the peak is due to the spin Zeeman effect since the valley-contrasting orbital moments in BLG only interact with the perpendicular magnetic field $B_\perp$. We focus on the s-shell first (see the upper panel). The first two peaks lean to the left as the magnetic field increases, while the 3$^{rd}$ and 4$^{th}$ peaks exhibit a different behavior. To quantify the peak shift, we use $(2\alpha/\mu_B)\,dV_{FG2}/dB_\parallel$ as an estimate of spin $g$ factor of each filled electron [46]. Here, $\alpha$ is the voltage-energy conversion coefficient, also known as the lever arm. As shown in Fig. 3(b), the 4 electrons in the s-shell can be divided into two groups, with negative/positive values corresponding to spin-up/spin-down, respectively. We also identify their valley indices by tracing the peak evolution under different $B_\perp$ (see Fig. 3(c)). These peak shifts under $B_\perp$ are mainly caused by the valley Zeeman effect since the valley $g$ factor is typically one order of magnitude larger than the spin $g$ factor in BLG [32,33]. The trend of valley pairing is observed, suggesting electrons with opposite valleys are filled sequentially. Therefore, the electron filling sequence can be summarized in Fig. 3(d). Two spin-up electrons from different valleys ($K$ and $K'$) are loaded in the dot first, followed by two spin-down electrons, in $K$ and $K'$ valleys respectively, to fulfill the s-shell. This indicates the two-electron ground state in BLG is a spin-triplet valley-singlet state, as revealed in previous studies [35,36]. Similarly, the electron filling sequence in the p-shell is also mapped, as shown in Fig. 3(d) (more details can be found in Section S6 in Supplemental Material [49]). Except for the difference that the p-shell has two orbitals, both s- and p-shells tend to be filled with electrons of the same polarization until half-filling. Also, the higher shell is occupied only after the lower shell is fully filled. These behaviors are consistent with the Hund's rules. We label the electron filling sequence here as "2+2+4+4".

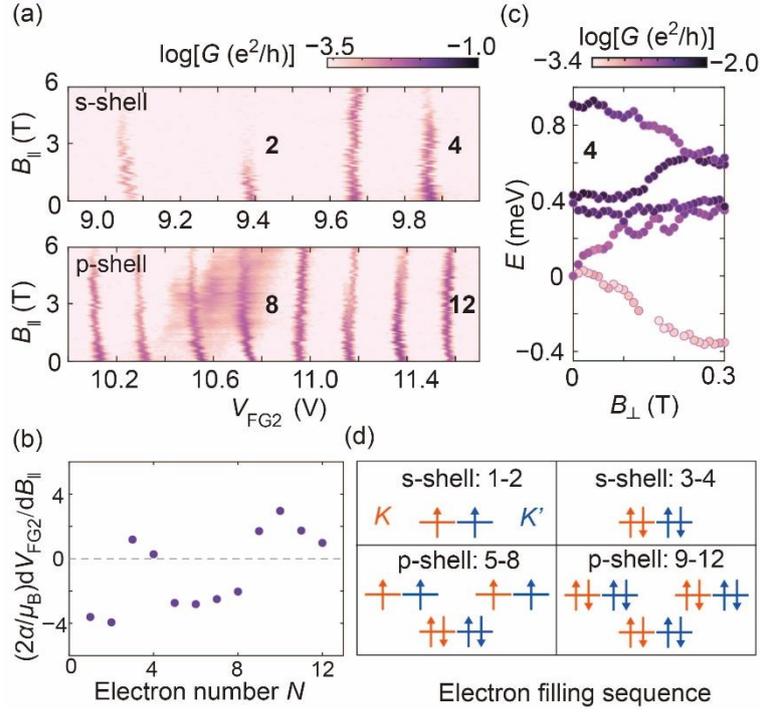

**Fig. 3** (a) Evolution of the first 12 Coulomb peaks under a parallel magnetic field $B_\parallel$ at a perpendicular electric field of $0.47\,\text{V/nm}$. (b) Fitted slopes $(2\alpha/\mu_B)\text{d}V_{\text{FG2}}/\text{d}B_\parallel$ of each Coulomb peaks from (a), giving an average value of 2.2. (c) Single-particle energy dispersion with perpendicular magnetic field $B_\perp$. The charging energy between neighboring peaks has been subtracted. (d) Schematic of spin/valley filling sequence of the first 12 electrons for the case of four-fold degeneracy.

As shown in Fig. 4, we map the spin/valley filling sequence at a large electric field of $0.65\,\text{V/nm}$ (corresponding to the case of twelve-fold degeneracy). Although the signature of valley pairing (see Fig. 4(c)) is similar to that at the small electric field, the spin filling sequence changes dramatically. As demonstrated in Figs. 4(a) and 4(b), the first 6 electrons are spin-up while the next 6 electrons ($7^{\text{th}}$ to $12^{\text{th}}$) are spin-down, exhibiting a structure of "6+6". Figure 4(d) summarizes the electron filling sequence, where the Hund's rules are still obeyed. Up to 6 spin-up electrons from different minivalleys of $K$ and $K'$ valleys are sequentially loaded until the s-shell is half-filled, followed by another 6 spin-down electrons to pair the unoccupied states. However, from the current measurement, we are not able to further distinguish the three minivalleys from each other.

We would like to emphasize that the electrically controlled switch in the spin filling sequence, from four-fold to twelve-fold degeneracy, has never been observed before. It is estimated that the orbital energy is ~0.5 meV, the short-range electron-electron interaction is ~0.4 meV, and the valley $g$-factor is ~30 in our device. These values indicate a relatively large dot [47], which is beneficial to achieve a lower charge density to access the minivalleys at the band edge. Further theoretical calculations suggest the trigonal warping effect is sufficiently pronounced upon increasing the perpendicular electric field in our device, so that the switch from four-fold to twelve-fold degeneracy can be triggered (see Section S7 in Supplemental Material [49]).

The observed "6+6" electron filling structure indicates that the minivalley indeed acts as a quantum degree of freedom, so that electrons with the same spin and valley but different minivalleys can be filled to the lowest s-shell, fulfilling the Pauli exclusion principle. This provides the possibility of generating high-spin states with total spin as large as $s_z = 3$, offering a promising platform to study the strong exchange interaction. In addition, such a three-fold minivalley degree of freedom provides an opportunity to realize flavor QDs to explore the SU(3) symmetry [48]. More interestingly, our results demonstrate that the trigonal warping induced minivalley degree of freedom can serve as a powerful handle to electrically access and manipulate the spin degree of freedom through their interplay in between. Besides, the capabilities of controlling the trigonal warping effect and mapping spin/valley sequence using QDs can be applied to investigate the exotic electronic phases recently discovered in the trigonally warped BLG [6,7,16] at the single-particle level.

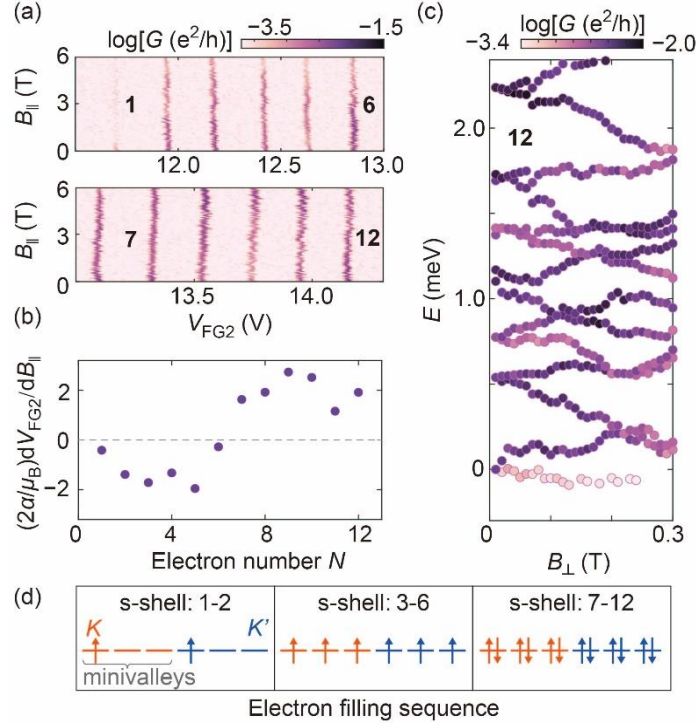

**Fig. 4** (a) Evolution of the first 12 Coulomb peaks under $B_\parallel$ at a perpendicular electric field of 0.65 V/nm. (b) Fitted slopes $(2\alpha/\mu_B)dV_{FG2}/dB_\parallel$ of each Coulomb peaks from (a), giving an average value of 1.6. (c) Single-particle energy dispersion with $B_\perp$. The charging energy between neighboring peaks has been subtracted. (d) Schematic of spin/valley filling sequence of the first 12 electrons for the case of twelve-fold degeneracy.

In conclusion, we investigate the shell filling structure of a BLG-based QD. Through electrical gating, we are able to control the three-fold minivalley induced by the trigonal warping to switch the degeneracy of the s-shell from four-fold to twelve-fold. Further magneto-measurements demonstrate the spin filling sequence is changed as well, which allows filling as much as 6, instead of 2, electrons of the same spin polarization to the s-shell in sequence. Our results reveal the possibility of manipulating the spin degree of freedom using the electrically-accessible minivalley degree of freedom.

# ACKNOWLEDGMENTS

The authors would like to thank Zhenhua Qiao and Qian Niu for fruitful discussions. This work was supported by the National Natural Science Foundation of China (Grant Nos. 12274397, 11904351, 12274401, and 12034018) and the Natural Science Foundation of Jiangsu Province (Grant No. BK20240123). This work was partially carried out at the USTC Center for Micro and Nanoscale Research and Fabrication.